\documentclass[english,3p]{elsarticle}
\usepackage[charter]{mathdesign}

\usepackage[T1]{fontenc}
\usepackage[utf8]{inputenc}
\usepackage{rotfloat}
\usepackage{booktabs}
\usepackage{amsmath}
\usepackage{graphicx}

\makeatletter

\providecommand{\tabularnewline}{\\}

\usepackage{amsthm}
\journal{arXiv}
\mathchardef\mhyphen="2D

\@ifundefined{showcaptionsetup}{}{%
 \PassOptionsToPackage{caption=false}{subfig}}
\usepackage{subfig}
\makeatother

\usepackage{babel}
\begin{document}

\begin{frontmatter}{}

\title{Measuring Systematic Risk with Neural Network Factor Model}

\author[labela]{Jeonggyu Huh\corref{cor1}}

\address[labela]{ School of Computational Sciences, Korea Institute for Advanced
Study, Seoul 02455, Republic of Korea}

\cortext[cor1]{corresponding author.\\
 E-mail address: aifina2018@kias.re.kr}
\begin{abstract}
In this paper, we measure systematic risk with a new nonparametric
factor model, the neural network factor model. The suitable factors
for systematic risk can be naturally found by inserting daily returns
on a wide range of assets into the bottleneck network. The network-based
model does not stick to a probabilistic structure unlike parametric
factor models, and it does not need feature engineering because it
selects notable features by itself. In addition, we compare performance
between our model and the existing models using 20-year data of S\&P
100 components. Although the new model can not outperform the best
ones among the parametric factor models due to limitations of the
variational inference, the estimation method used for this study,
it is still noteworthy in that it achieves the performance as best
the comparable models could without any prior knowledge. 
\end{abstract}
\begin{keyword}
systematic risk; neural network; nonparametric model; Bayesian statistics;
variational inference

JEL Classification: C6, C8, G01, G10
\end{keyword}

\end{frontmatter}{}

\section{Introduction}

Systematic risk is vulnerability to events which affect a great number
of assets or the whole market, in contrast with idiosyncratic risk
to a specific company. For example, the events such as bank failures
can be sources of the risk. In finance, systematic risk is considered
to be significant because even a well-diversified portfolio can not
hedge the risk (so it is often called undiversifiable risk). This
implies that identifying intrinsic systematic risk beyond asset returns
is inevitable for asset pricing and portfolio optimization (\citet{cochrane2009asset,das2004systemic}).
Apart from those, reliably measuring the risk is required in various
other areas such as equity option pricing (\citet{christoffersen2017factor}),
credit derivative pricing (\citet{duffie2009frailty}) and risk management
(\citet{kalkbrener2015correlation}).

Measuring systematic risk is generally done by one or more common
factors. These works are broadly divided into two directions, depending
on whether common factors are observable or not. Let us compare two
renowned asset pricing theories, the capital asset pricing model (CAPM)
and arbitrage pricing theory (APT) (\citet{sharpe1964capital,ross2013arbitrage}).
The factor for the CAPM should be the historical returns of a market
portfolio. In contrast, the APT does not have such a restriction,
so its factors are allowed to be found by mathematical algorithms
like factor analysis (\citet{roll1980empirical,chen1983some}). On
the other hand, various extensions of both the models have been proposed
by the influence of a great many studies regarding heteroskedasticity
and sudden jumps in single asset returns (\citet{chesney1989pricing,heston1993closed,duffie2000transform}).
For example, \citet{bentzen2003intertemporal} and \citet{ho1996continuous}
devised a CAPM reflecting jump risk and an APT with both stochastic
volatility and jumps, respectively. Moreover, several factor models,
which are originated for other purposes, can be also regarded as extensions
of the APT (\citet{jacquier1995models,Ray2000}).

Neural network theory can give better solutions to measuring systematic
risk than the parametric factor models holding fixed views regardless
of market characteristics. This is because the parameters and the
latent variables of a network can be learned in a manner dependent
on market conditions. To the best of author's knowledge, previous
approaches using neural networks for systematic risk mainly focus
on two subtly different problems. One is to study the exposure of
individual asset prices to systematic risk, for example, successful
estimation or prediction of the beta for the CAPM (\citet{wittkemper1996using,yuan2015using}).
The other is to reflect on the risk itself such as accurate predictions
of financial crises (\citet{celik2007evaluating,dabrowski2016systemic}).
However, it seems that these methods can be improved by removing ad
hoc properties or feature engineering from them. The beta prediction
might be reckoned a make-do in that the beta should not change in
principle. In fact, a rigorous method is to extend the CAPM itself
rather than to be reliant too much on market-implied betas. Furthermore,
it is difficult for a researcher to choose features without subjective
judgements. If possible, one may as well automate the process.

In this paper, we measure systematic risk with a new nonparametric
factor model, the neural network factor model (NNFM). The NNFM is
strongly motivated by the variational recurrent neural network (VRNN)
of \citet{chung2015recurrent}. Simply speaking, the VRNN is a generative
model to put a Bayesian perspective into a recurrent neural network
(RNN), thereby resulting in an effective description of data sequence
with high variability. In a common design of the NNFM, the number
of the nodes in the deepest layer is much smaller than the number
of input nodes. The structure is intended to make the deepest layer
a bottleneck in data transmission and leave essential features of
the given data to the layer. Pondering on an assessment method for
systematic risk with the NNFM, we come up with the idea that suitable
factors for systematic risk can be naturally found by inserting daily
returns on a wide range of assets into the bottleneck network. This
approach has several advantages compared with the parametric factor
models such as the APT and the previous network-based methods. First,
it does not stick to a probabilistic structure unlike the parametric
models. Second, it does not employ any ad hoc approach like the implied
beta. Finally, it does not need feature engineering because the NNFM
selects notable features by itself.

We compare performance between the NNFM and the parametric factor
models for measuring systematic risk based on two criteria: marginal
log-likelihood and variational lower bound, which have been widely
utilized for model selection in Bayesian statistics (c.f. \citet{chung2015recurrent,kingma2015variational}).
For this end, we use 5,240-day data of S\&P 100 components from 16
May 1997 to 13 March 2018 (about 20 years). The parameters and the
latent variables of the models are estimated using a Bayesian method
called variational inference. We find that the NNFM runs into the
overfitting problem, but it is resolved by inventing a modification
of the NNFM named as the monotone NNFM. However, we can not verify
that the network-based models outperform the best ones among the parametric
factor models. This is because we are unable to elicit the maximum
capacities of the models since the variational inference seeks approximate
posteriors of systematic factors. But the new models are still noteworthy
in that they do not require any prior knowledge. A researcher using
the models need not worry about which one is a proper choice among
the existing models. The network-based models can achieve the performance
as best the parametric factor models could. 

In the next section, we introduce the NNFM and review the parametric
factor models. In Section \ref{sec:variational_inference}, we explain
an estimation method of the NNFM using the variational inference and
sum up all the discussions theretofore. Section \ref{sec:empirical_analysis}
shows an empirical test and its results. The last section concludes.

\section{Neural Network Factor Model}

\subsection{General Factor Model\label{subsec:GFM}}

In this paper, the return process $\boldsymbol{x}_{t}=(x_{t,1},x_{t,2},\cdots,x_{t,n_{x}})$
of $n_{x}$-assets is assumed to be randomly generated by a model
involving a $n_{z}$-dimensional factor process $\boldsymbol{z}_{t}=(z_{t,1},z_{t,2},\cdots,z_{t,n_{z}})$.
The model is expressed by 
\begin{gather}
\boldsymbol{x}_{t}|\boldsymbol{z}_{t}\sim\boldsymbol{N}\left(\boldsymbol{\mu}_{x|z,t},\boldsymbol{\sigma}_{x|z,t}^{2}\boldsymbol{I}_{n_{X}}\right),\;{\rm where}\;[\boldsymbol{\mu}_{x|z,t},\boldsymbol{\sigma}_{x|z,t}^{2}]=\varphi_{x|z}\left(\boldsymbol{z}_{t};\Theta_{x|z}\right),\label{eq:generative_x}\\
\boldsymbol{z}_{t}|\boldsymbol{z}_{<t}\sim\boldsymbol{N}\left(\boldsymbol{\mu}_{z,t},\boldsymbol{\sigma}_{z,t}^{2}\boldsymbol{I}_{n_{z}}\right),\;{\rm where}\;[\boldsymbol{\mu}_{z,t},\boldsymbol{\sigma}_{z,t}^{2}]=\varphi_{z}\left(\boldsymbol{z}_{<t};\Theta_{z}\right),\label{eq:generative_z}
\end{gather}
where $t=1,2,\cdots,T$, $\boldsymbol{N}$ is a multivariate normal
distribution, $\boldsymbol{I}_{n_{x}}$ and $\boldsymbol{I}_{n_{z}}$
are the identity matrices of sizes $n_{x}$ and $n_{z}$, respectively,
$(\boldsymbol{\mu}_{x|z,t},\boldsymbol{\sigma}_{x|z,t}^{2})$ and
$(\boldsymbol{\mu}_{z,t},\boldsymbol{\sigma}_{z,t}^{2})$ control
the means and the variances of $\boldsymbol{x}_{t}|\boldsymbol{z}_{t}$
and $\boldsymbol{z}_{t}$, respectively, $\varphi_{x|z}:\mathbb{R}^{n_{z}}\rightarrow\mathbb{R}^{n_{x}}\times\mathbb{R}^{n_{x}}$,
$\varphi_{z}:\mathbb{R}^{n_{z}\times t}\rightarrow\mathbb{R}^{n_{z}}\times\mathbb{R}^{n_{z}}$,
$\Theta_{x|z}$ and $\Theta_{z}$ are the parameter sets to determine
the structures of $\varphi_{x|z}$ and $\varphi_{z}$, respectively,
$\boldsymbol{z}_{<t}:=\{\boldsymbol{z}_{0},\boldsymbol{z}_{1},\boldsymbol{z}_{2},\cdots,\boldsymbol{z}_{t-1}\}$,
and $\boldsymbol{z}_{0}$ is assumed to be known. The dimension of
$\boldsymbol{z}_{t}$ is much smaller than the number of the assets,
i.e. $n_{z}\ll n_{x}$, so that $\boldsymbol{z}_{t}$ represents systematic
factors for the given assets. As remarked in the introduction, factor
approaches have been widely preferred as research tools for systematic
risk. On the other hand, note that the proposed model can be considered
a non-Markov model as well as a Markov model because $\boldsymbol{z}_{t}$
may depend on its all past values $\boldsymbol{z}_{<t}$. We call
this model the general factor model (GFM).

If necessary, the GFM can be specified by giving explicit forms to
$\varphi_{x|z}$ and $\varphi_{z}$, which implies that it embraces
various models. We pick a simple example corresponding to the case
that $n_{x}=2$ and $n_{z}=1$:
\begin{gather*}
x_{t,i}=\alpha_{0,i}+\sqrt{\beta_{0,i}+\beta_{1,i}z_{t}}\epsilon_{t,i},\\
z_{t}=c+az_{t-1}+\sqrt{z_{t-1}}e_{t},
\end{gather*}
where $i=1,2$, and $\epsilon_{t,i}$ and $e_{t}$ are independent
standard normal variables. This model can be easily converted into
a consistent form with the GFM as follows:
\begin{gather*}
\boldsymbol{x}_{t}|z_{t}\sim\boldsymbol{N}\left(\boldsymbol{\mu}_{x|z},\boldsymbol{\Sigma}_{x|z}\right),\\
z_{t}\sim N\left(c+az_{t-1},z_{t-1}\right),
\end{gather*}
where
\[
\boldsymbol{x}_{t}=\left[\begin{array}{c}
x_{t,1}\\
x_{t,2}
\end{array}\right]\;,\boldsymbol{\mu}_{x|z}=\left[\begin{array}{c}
\alpha_{0,1}\\
\alpha_{0,2}
\end{array}\right],\;\boldsymbol{\Sigma}_{x|z}=\left[\begin{array}{cc}
\beta_{0,1}+\beta_{1,1}z_{t-1} & 0\\
0 & \beta_{0,2}+\beta_{1,2}z_{t-1}
\end{array}\right].
\]
Note that $\Theta_{x|z}=\{\alpha_{0,1},\alpha_{0,2},\beta_{0,1},\beta_{0,2},\beta_{1,1},\beta_{1,2}\}$
and $\Theta_{z}=\{c,a\}$ for this example.

\subsection{Existing Factor Models\label{subsec:existing_models}}

We shortly review some existing factor models: three parametric ones
(APT, L-SVFM, SR-SVFM), a nonparametric one (G-SVFM) and two hybrid
ones integrating the parametric models (APT-L, APT-SR). Later, the
parametric factor models and the hybrid models among them will be
compared with our network-based approach – the nonparametric model
is presented here to show its weak point. In all the below models,
every parameter except $\alpha_{0,i}$ and $\beta_{0,i}$ is positive,
and $\epsilon_{t,i}$, $e_{t,j}$ are independent standard normal
variables, $i=1,2,\cdots,n_{x}$($n_{x}\gg2$) and $j=1,2$. For a
brief notation, only the two factor case, i.e. $n_{z}=2$, for each
model is illustrated, which can be easily generalized if necessary.
\begin{itemize}
\item Arbitrage pricing theory (APT)
\begin{gather*}
x_{t,i}=\alpha_{0,i}+\alpha_{1,i}z_{t,1}+\alpha_{2,i}z_{t,2}+\beta_{0,i}\epsilon_{t,i},\\
z_{t,j}=e_{t,j}.
\end{gather*}
In the APT, the expected asset return is modeled as a linear sum of
factor processes (\citet{roll1980empirical}). The processes can be
macroeconomic ones such as surprises in inflation, or inexplicable
ones to be found by mathematical algorithms. We choose the latter
way and infer the factor processes from data. The parameters for the
APT are commonly obtained by factor analysis, but we estimate them
using a Bayesian method, variational inference, for consistent comparisons
with other models. If $n_{z}=1$ and $z_{t,1}$ is set as the return
process of a market portfolio, the APT is reduced to the CAPM. 
\item Logarithmic stochastic volatility factor model (L-SVFM)
\begin{gather*}
x_{t,i}=\alpha_{0,i}+\exp\left(\beta_{0,i}+\beta_{1,i}z_{t,1}+\beta_{2,i}z_{t,2}\right)\epsilon_{t,i},\\
z_{t,j}=a_{j}z_{t-1,j}+e_{t,j},
\end{gather*}
where $z_{t,j}$ is a stationary process ($0<a_{j}<1$) reverting
to zero. The L-SVFM is a slight modification of the factor model in
\citet{Ray2000}, which falls into the category of stochastic volatility
factor models (SVFM). As widely known, stochastic volatility can account
for several stylized facts such as clustering and mean-reversion of
volatility, which are challenging for the APT to reflect. According
to \citet{chernov2003alternative} and \citet{raggi2006comparing},
it may be empirically expected that the L-SVFM is better than the
SR-SVFM below in generating realistic return behavior.
\item Square-root stochastic volatility factor model (SR-SVFM)
\begin{gather*}
x_{t,i}=\alpha_{0,i}+{\rm sqrt}\left(\beta_{0,i}+\beta_{1,i}z_{t,1}+\beta_{2,i}z_{t,2}\right)\epsilon_{t,i},\\
z_{t,j}=c_{j}+a_{j}z_{t-1,j}+\sqrt{z_{t-1,j}}e_{t,j},
\end{gather*}
where ${\rm sqrt}\left(x\right)=\sqrt{x}$, $z_{t,j}$ is a stationary
process ($0<a_{j}<1$) reverting to $c_{j}$ and satisfies the Feller
condition for the sake of positivity. We adopt the SR-SVFM, which
looks like a natural extension of \citet{heston1993closed}, among
the affine diffusion models of \citet{duffie2000transform}. The affine
structure of the SR-SVFM may not be suitable to describing actual
asset dynamics (\citet{jones2003dynamics}). However, these types
of models often provide analytic pricing formulas for derivatives,
so they are preferred as quick ways to price the products.
\item Generalized stochastic volatility factor model (G-SVFM)
\begin{gather*}
x_{t,i}=\alpha_{0,i}+\sum_{p,q}d_{p,q}H_{p}\left(z_{t,1}\right)H_{q}\left(z_{t,2}\right)\epsilon_{t,i},\\
z_{t,j}=a_{j}z_{t-1,j}+e_{t,j},
\end{gather*}
or
\begin{gather*}
x_{t,i}=\alpha_{0,i}+\sum_{p,q}d_{p,q}L_{p}\left(z_{t,1}\right)L_{q}\left(z_{t,2}\right)\epsilon_{t,i},\\
z_{t,j}=c_{j}+a_{j}z_{t-1,j}+\sqrt{z_{t-1,j}}e_{t,j},
\end{gather*}
where $H$ and $L$ are the Hermite and Laguerre polynomials, respectively.
First, note that $z_{t,j}$ in the L-SVFM and the SR-SVFM can be considered
as Euler's discretizations of the Ornstein–Uhlenbeck process and the
Cox–Ingersoll–Ross process, respectively. The polynomials $H$ and
$L$ are the eigenfunctions for the infinitesimal operators of the
processes (cf. \citet{fouque2011multiscale}). It means that these
models can be accepted as generalizations of the L-SVFM and the SR-SVFM.
Refer to \citet{meddahi2001eigenfunction} for more details. This
model is the most well-known among only a few nonparametric factor
models but has a critical weak point that $\boldsymbol{z}_{t}$ should
belong to the domain to ensure positive volatility of $x_{t,i}$ during
the estimation process. On the contrary, our nonparametric model is
designed to be free from such a constraint. 
\item Hybrid model (APT-L and APT-SR)
\begin{gather*}
x_{t,i}=\alpha_{0,i}+\alpha_{1,i}z_{t,1}^{\left(1\right)}+\alpha_{2,i}z_{t,2}^{{\rm \left(1\right)}}+f\left(z_{t,1}^{\left(2\right)},z_{t,2}^{\left(2\right)}\right)\epsilon_{t,i},\\
z_{t,j}^{\left(1\right)}=e_{t,j},\quad z_{t,k}^{\left(2\right)}=g\left(z_{t-1,k}^{\left(2\right)},e_{t,k}\right),
\end{gather*}
where $j=1,2$ and $k=1,2$. The hybrid model represents a model integrating
the APT and one kind of the SVFM. Considering only the aforementioned
models, the possible hybrid models are the APT-L (the APT + the L-SVFM)
and the APT-SR (the APT + the SR-SVFM). In other words, $f$ is one
of the two functions, $\exp(u(\cdot,\cdot))$ and ${\rm sqrt}(u(\cdot,\cdot))$,
where $u(x,y)=\beta_{0,i}+\beta_{1,i}x+\beta_{2,i}y$. $g$ should
be also determined in accordance with $f$. For example, for the APT-L,
$g$ must be a function which makes $z_{t}$ an AR(1) process.
\end{itemize}
One needs to know that the GFM includes all the above models. Namely,
each model can be viewed as a specification of the GFM.

\subsection{Neural Network Factor Model\label{subsec:NNFM}}

\citet{chung2015recurrent} extended the well-known variational autoencoder
of \citet{kingma2013auto} and invented the variational recurrent
neural network (VRNN), which is a generative model to put a Bayesian
perspective into recurrent neural networks (RNN). The VRNN is theoretically
appealing, but it seems improper for modeling asset returns because
it does not consider the efficient-market hypothesis. So, a naive
application of the VRNN leads to a market view inconsistent with most
financial models. To modify the VRNN for our needs, we now add a new
nonparametric model in the class of the GFM, the neural network factor
model (NNFM). 

Non-Markov models are flexible but difficult to estimate. Recall that
the GFM can be non-Markov. So, in order to remove the non-Markov property
from the GFM with the least loss of past memories, we first introduce
$\boldsymbol{h}_{t}=(h_{t,1},h_{t,2},\cdots,h_{t,n_{h}})$ to store
the past and present values of $\boldsymbol{z}_{t}$. Every time $\boldsymbol{z}_{t}$
is changed, the storage $\boldsymbol{h}_{t}$ is updated through the
following recurrence relation: 
\begin{equation}
\boldsymbol{h}_{t}=\varphi_{h}\left(\boldsymbol{z}_{t},\boldsymbol{h}_{t-1};\Theta_{h}\right),\label{eq:recurrence}
\end{equation}
where $\Theta_{h}$ is the parameter set to determine the structure
of $\varphi_{h}:\mathbb{R}^{n_{z}}\times\mathbb{R}^{n_{h}}\rightarrow\mathbb{R}^{n_{h}}$,
and $\boldsymbol{h}_{0}$ is assumed to be known. If $\left(\boldsymbol{h}_{t},\varphi_{h}\right)$
is built with enough consideration, there is a continuous function
$\psi$ such that $\psi\left(\boldsymbol{h}_{t}\right)=\boldsymbol{z}_{\leq t}(:=\left\{ \boldsymbol{z}_{0},\boldsymbol{z}_{1},\cdots,\boldsymbol{z}_{t}\right\} )$
almost holds under some measures. In other words, $\psi$ can faithfully
reproduce $\boldsymbol{z}_{\leq t}$ with $\boldsymbol{h}_{t}$. If
$\varphi_{h}$ is constructed using feedforward neural networks (FNN),
this claim can be supported by the universal approximation theorem
for the RNN (\citet{funahashi1993approximation}). Then, by putting
$\varphi_{z}^{*}\left(\boldsymbol{h}_{t};\Theta_{z}\right)=\varphi_{z}\left(\psi\left(\boldsymbol{h}_{t}\right);\Theta_{z}\right)$,
the generative rule (\ref{eq:generative_z}) on $\boldsymbol{z}_{t}$
is expressed as follows: 
\begin{gather}
\boldsymbol{z}_{t}\sim\boldsymbol{N}\left(\boldsymbol{\mu}_{z,t}^{*},\boldsymbol{\sigma}_{z,t}^{*2}\boldsymbol{I}_{n_{z}}\right),\;{\rm where}\;[\boldsymbol{\mu}_{z,t}^{*},\boldsymbol{\sigma}_{z,t}^{*2}]=\varphi_{z}^{*}\left(\boldsymbol{h}_{t-1};\Theta_{z}\right).\label{eq:generative_z_2}
\end{gather}

Because we aim to take a network-based approach, the functions $\varphi_{x}$
in (\ref{eq:generative_x}), $\varphi_{z}^{*}$ in (\ref{eq:generative_z_2})
and $\varphi_{h}$ in (\ref{eq:recurrence}) are implemented through
their respective FNNs:
\[
\varphi_{x|z}\left(\boldsymbol{z}_{t};\Theta_{x|z}\right)={\rm FNN}_{x|z}\left(\boldsymbol{z}_{t};\Theta_{x|z}\right),\;\varphi_{z}^{*}\left(\boldsymbol{h}_{t-1};\Theta_{z}\right)={\rm FNN}_{z}\left(\boldsymbol{h}_{t-1};\Theta_{z}\right),\;\varphi_{h}\left(\boldsymbol{z}_{t},\boldsymbol{h}_{t-1}\right)={\rm FNN}_{h}\left(\boldsymbol{z}_{t},\boldsymbol{h}_{t-1};\Theta_{h}\right).
\]
Continuous functions $\varphi_{x|z}$, $\varphi_{z}^{*}$ and $\varphi_{h}$
can be well approximated by the FNNs with enough number of hidden
nodes (\citet{cybenko1989approximation,hornik1991approximation}).
By virtue of the universal approximation theorems for the FNN and
the RNN, we expect that the NNFM may be the maximal submodel of the
GFM which possesses all specifications of the GFM. Moreover, the storage
capacity of ($\boldsymbol{h}_{t},\varphi_{h}$) can be improved using
more advanced methods such as long short-term memory unit (LSTM) and
gated recurrent unit (GRU). We use LSTM, more prevalent than the others,
for the NNFM. For further details on these networks, refer to the
relevant chapters in \citet{goodfellow2016deep}.

The NNFM has a drawback that meanings of $\boldsymbol{z}_{t}$ are
rather unclear. So, we additionally propose a submodel of the NNFM,
the M-NNFN (monotone NNFM), which comprises the M-NNFM(1) and the
M-NNFM(2). First, let us examine the following network, the MI-FNN
(monotone increasing FNN):
\begin{gather*}
{\rm MI\mhyphen FNN}_{x|z,\mu}\left(\boldsymbol{z};\Theta_{x|z}\right)={\rm PReLU}\left({\rm PReLU}\left(\boldsymbol{z}\exp\left(W_{1,\mu}\right)+b_{1,\mu}\right)\exp\left(W_{2,\mu}\right)+b_{2,\mu}\right),\\
{\rm MI\mhyphen FNN}_{x|z,\sigma}\left(\boldsymbol{z};\Theta_{x|z}\right)={\rm SoftPlus}\left({\rm PReLU}\left(\boldsymbol{z}\exp\left(W_{1,\sigma}\right)+b_{1,\sigma}\right)\exp\left(W_{2,\sigma}\right)+b_{2,\sigma}\right),
\end{gather*}
where ${\rm PReLU}(x)=xI_{x\geq0}+0.5xI_{x<0}$, ${\rm SoftPlus}\left(x\right)=\log\left(1+e^{x}\right)$,
$W_{1,\mu}$ and $W_{1,\sigma}$ are $n_{z}\times m$ matrices, $W_{2,\mu}$
and $W_{2,\sigma}$ are $m\times n_{x}$ matrices, $b_{1,\mu}$ and
$b_{1,\sigma}$ are $m$-dimensional vectors, $b_{2,\mu}$ and $b_{2,\sigma}$
are $n_{x}$-dimensional vectors, $m$ is the number of the nodes
in the hidden layer, and all the functions can be applied elementwise
to matrices and vectors. Note that $\Theta_{x|z}=\{W_{1,\mu},W_{2,\mu},b_{1,\mu},b_{2,\mu},W_{1,\sigma},W_{2,\sigma},b_{1,\sigma},b_{2,\sigma}\}$
for the MI-FNN. One can check without much effort that the MI-FNN
is monotonically increasing with respect to $\boldsymbol{z}$. 

We now construct the M-NNFM(1) and the M-NNFM(2) with the MI-FNN.
For the M-NNFM(1), $\boldsymbol{z}_{t}$ is one-dimensional, i.e.
$\boldsymbol{z}_{t}=(z_{t,1})$, and $\boldsymbol{\mu}_{x|z,t}$ and
$\boldsymbol{\sigma}_{x|z,t}^{2}$ are monotonically decreasing and
increasing with respect to $z_{t,1}$, respectively, that is to say,
\[
\boldsymbol{\mu}_{x|z,t}=-{\rm MI\mhyphen FNN}_{x|z,\mu}\left(z_{t,1};\Theta_{x|z}\right),\;\boldsymbol{\sigma}_{x|z,t}^{2}={\rm MI\mhyphen FNN}_{x|z,\sigma}\left(z_{t,1};\Theta_{x|z}\right).
\]
This model is designed to accommodate the leverage effect, the inverse
relationship between asset prices and volatility. The M-NNFM(1) makes
it possible for systematic risk to be assessed by the only one factor
$z_{t,1}$; $z_{t,1}$ would be high in turbulent markets but low
in normal markets, which means that $z_{t,1}$ can be utilized as
an economic indicator. On the other hand, as for the M-NNFM(2), $\boldsymbol{z}_{t}$
is two-dimensional, i.e. $\boldsymbol{z}_{t}=(z_{t,1},z_{t,2})$,
and $\boldsymbol{\mu}_{x|z,t}$ and $\boldsymbol{\sigma}_{x|z,t}^{2}$
are monotone increasing functions of $z_{t,1}$ and $z_{t,2}$, respectively,
as below:
\[
\boldsymbol{\mu}_{x|z,t}={\rm MI\mhyphen FNN}_{x|z,\mu}\left(z_{t,1};\Theta_{x|z}\right),\;\boldsymbol{\sigma}_{x|z,t}^{2}={\rm MI\mhyphen FNN}_{x|z,\sigma}\left(z_{t,2};\Theta_{x|z}\right),
\]
This design helps us to consider $z_{t,1}$ and $z_{t,2}$ the respective
indices for $\boldsymbol{\mu}_{x|z,t}$ and $\boldsymbol{\sigma}_{x|z,t}^{2}$;
$z_{t,1}$ would be high in bull markets but low in bear markets,
and $z_{t,2}$ would be high in volatile markets but low in steady
markets. Along with $z_{t,1}$ in the M-NNFM(1), $(z_{t,1},z_{t,2})$
can be also used as another economic indicator. Moreover, one can
measure the sensitivity of each asset to systematic risk by observing
the differentials $\partial_{z_{j}}\boldsymbol{\mu}_{x|z,t}$ and
$\partial_{z_{j}}\boldsymbol{\sigma}_{x|z,t}$. This can be a great
help in doing a thorough management of systematic risk.

\section{\label{sec:variational_inference}Variational Inference Neural Network}

\subsection{Variational Inference\label{subsec:VI}}

Following the spirit of Bayesians, we infer the posterior $p(\boldsymbol{z}_{\leq T}|\boldsymbol{x}_{\leq T};\Theta)$
of the factor process $\boldsymbol{z}_{\leq T}$ for the GFM, where
$\Theta$ means all the parameters $\Theta_{x|z}\cup\Theta_{z}\cup\Theta_{h}$.
At a glance, it seems possible that the posterior can be calculated
by Bayes' theorem as follows:
\[
p\left(\boldsymbol{z}_{\leq T}|\boldsymbol{x}_{\leq T};\Theta\right)=\frac{p\left(\boldsymbol{x}_{\leq T}|\boldsymbol{z}_{\leq T};\Theta\right)p\left(\boldsymbol{z}_{\leq T};\Theta\right)}{p\left(\boldsymbol{x}_{\leq T};\Theta\right)}.
\]
However, the marginal likelihood $p(\boldsymbol{x}_{\leq T};\Theta)$
is analytically intractable for most specifications of the GFM in
contrast with the likelihood $p(\boldsymbol{x}_{\leq T}|\boldsymbol{z}_{\leq T};\Theta)$
and the prior $p(\boldsymbol{z}_{\leq T};\Theta)$ which can be induced
from (\ref{eq:generative_x}) and (\ref{eq:generative_z}). Thus,
numerical techniques must be used such as the Markov chain Monte-Carlo
simulation (MCMC) and the variational inference (VI). In this paper,
the VI is only employed because MCMC computations for complex models
are so time-consuming; the GFM has tens of thousands of latent variables
for modeling long-term data of many assets. Refer to \citet{kruschke2014doing}
for a comprehensive review of Bayesian statistics.

The VI overcomes this difficulty by giving a tractable approximation
$q(\boldsymbol{z}_{\leq T}|\boldsymbol{x}_{\leq T};\phi)$ for $p(\boldsymbol{z}_{\leq T}|\boldsymbol{x}_{\leq T};\Theta)$,
where $\phi$ is the parameter set for $q$. The approximation brings
a considerable reduction in computational time compared with the MCMC.
Of course, it is desirable to minimize a distance, such as the Kullback–Leibler
(KL) divergence, between $q(\boldsymbol{z}_{\leq T}|\boldsymbol{x}_{\leq T};\phi)$
and $p(\boldsymbol{z}_{\leq T}|\boldsymbol{x}_{\leq T};\Theta)$.
The goal can be achieved by maximizing the variational lower bound
$\mathcal{L}(\Theta,\phi;\boldsymbol{x}_{\leq T})$ with respect to
$\Theta$ and $\phi$. $\mathcal{L}$ can be given by
\begin{align}
\mathcal{L}\left(\Theta,\phi;\boldsymbol{x}_{\leq T}\right) & =\frac{1}{L}\sum_{l=1}^{L}\log p_{q\left(\phi\right)}\left(\boldsymbol{x}_{\leq T}|\boldsymbol{z}_{\leq T}^{\left(l\right)};\Theta\right)-{\rm KL}\left[\left.q\left(\boldsymbol{z}_{\leq T}|\boldsymbol{x}_{\leq T};\phi\right)\right\Vert p\left(\boldsymbol{z}_{\leq T};\Theta\right)\right]\nonumber \\
 & =\sum_{t=1}^{T}\left(\frac{1}{L}\sum_{l=1}^{L}\log p_{q\left(\phi\right)}\left(\boldsymbol{x}_{t}|\boldsymbol{z}_{t}^{\left(l\right)};\Theta\right)-{\rm KL}\left[\left.q\left(\boldsymbol{z}_{t}|\boldsymbol{x}_{t},\boldsymbol{z}_{<t};\phi\right)\right\Vert p\left(\boldsymbol{z}_{t}|\boldsymbol{z}_{<t};\Theta\right)\right]\right),\label{eq:variational_lower_bound}
\end{align}
where the subscript $q(\phi)$ clarifies that $\boldsymbol{z}_{\leq T}^{\left(l\right)}$
is sampled from the distribution $q$, and $L$ is the number of simulations
of $\boldsymbol{z}_{\leq T}$. The last equality follows from $p(\boldsymbol{z}_{\leq T}|\boldsymbol{x}_{\leq T};\Theta)=\Pi_{t=1}^{T}p(\boldsymbol{z}_{t}|\boldsymbol{x}_{t,}\boldsymbol{z}_{<t};\Theta)$
and the additivity of the KL divergence. The first term for the above
formula is the expected log-likelihood to gauge the degree of reconstruction.
The second term is a regularizer to measure how much information is
lost when using the approximate posterior. It is also notable that
the KL divergence is often computed analytically. Even if not, it
can be quickly obtained using numerical schemes. 

The VI for the GFM is, in short, to maximize $\mathcal{L}(\Theta,\phi;\boldsymbol{x}_{\leq T})$
with respect to $\Theta$ and $\phi$, thereby achieving the approximate
posterior $q(\boldsymbol{z}_{\leq T}|\boldsymbol{x}_{\leq T};\phi)$
along with the parameters $\Theta$. Let us explain how the optimization
proceeds. First, $\boldsymbol{z}_{\leq T}$ is simulated for given
$(\Theta,\phi)$. Second, holding the generated factor process $\boldsymbol{z}_{t}$,
$(\Theta,\phi)$ is updated through a numerical optimization such
as the stochastic gradient descent method. The two processes are repeated
until convergence. A useful tip is that $L$ can be set to be $1$
if $T$ is large enough. This makes the implementation of the optimization
much easier. In addition, note that the starting value of $\boldsymbol{z}_{t}$,
i.e. $\boldsymbol{z}_{0}$, should be set for the optimization. One
could worry about the turbulence caused by the arbitrary setting.
But, as the expression (\ref{eq:variational_lower_bound}) hints,
quite a large $T$ can make fitting results robust to the initial
setting. In other words, one only needs to use big data in order to
leave no room for subjective judgement. 

Lastly, we emphasize that the VI for this work is not fully Bayesian
as our argument has not involved the notion ``hyperpriors'' for
the parameters $\Theta$. In other words, $\Theta$ are not random
variables but deterministic variables. In fact, according to some
recent papers, there is a strong possibility that the full VI assuming
stochastic parameters shows higher performance. However, we have to
implement the latest advance called variational dropout so as to realize
the full VI for network-based approaches (cf. \citet{kingma2015variational,gal2016theoretically}).
Because the work seems to be out of scope for this paper, we will
investigate this topic in further studies.

\subsection{Variational Inference Neural Network\label{subsec:VINN}}

The NNFM is a generative network to generate $(\boldsymbol{x}_{t},\boldsymbol{h}_{t})$
from $(\boldsymbol{z}_{t},\boldsymbol{h}_{t-1})$. Conversely, the
variational inference neural network (VINN) is an inference network
to infer $(\boldsymbol{z}_{t},\boldsymbol{h}_{t})$ from $(\boldsymbol{x}_{t},\boldsymbol{h}_{t-1})$.
As the counterpart of the NNFM, the VINN aims to implement $q^{*}(\boldsymbol{z}_{t}|\boldsymbol{x}_{t},\boldsymbol{h}_{t-1};\phi)$
with a neural network, where $q^{*}(\boldsymbol{z}_{t}|\boldsymbol{x}_{t},\boldsymbol{h}_{t-1};\phi):=q(\boldsymbol{z}_{t}|\boldsymbol{x}_{t},\psi\left(\boldsymbol{h}_{t-1}\right);\phi)$. 

For the VINN, $q^{*}(\boldsymbol{z}_{t}|\boldsymbol{x}_{t},\boldsymbol{h}_{t-1};\phi)$
is set as a conditional multivariate normal distribution
\begin{gather}
\boldsymbol{z}_{t}|\boldsymbol{x}_{t}\sim\boldsymbol{N}\left(\boldsymbol{\mu}_{z|x,t},\boldsymbol{\sigma}_{z|x,t}^{2}\boldsymbol{I}_{n_{z}}\right),\;{\rm where}\;[\boldsymbol{\mu}_{z|x,t},\boldsymbol{\sigma}_{z|x,t}^{2}]=\varphi_{z|x}^{*}\left(\boldsymbol{x}_{t},\boldsymbol{h}_{t-1};\phi\right),\label{eq:infer_z}
\end{gather}
where $(\boldsymbol{\mu}_{z|x,t},\boldsymbol{\sigma}_{z|x,t}^{2})$
controls the mean and the variance of $\boldsymbol{z}_{t}|\boldsymbol{x}_{t}$,
and $\varphi_{z|x}^{*}:\mathbb{R}^{n_{x}}\times\mathbb{R}^{n_{h}}\rightarrow\mathbb{R}^{n_{z}}\times\mathbb{R}^{n_{z}}$.
Then, because the KL divergence between two normal distributions are
analytically known, the second term in (\ref{eq:variational_lower_bound})
needs not be computed numerically. As in the NNFM, $\varphi_{z|x}^{*}$is
also specified by a FNN:
\begin{equation}
\varphi_{z|x}^{*}\left(\boldsymbol{x}_{t},\boldsymbol{h}_{t-1};\phi\right)={\rm FNN}_{z|x}\left(\boldsymbol{x}_{t},\boldsymbol{h}_{t-1};\phi\right).\label{eq:varphi_z|x}
\end{equation}
The VINN can be also used for estimating the parametric factor models
in Subsection \ref{subsec:existing_models}. Recall that the models
are all Markov, i.e., $\boldsymbol{z}_{t}$ only depends on $\boldsymbol{z}_{t-1}$.
So, $\boldsymbol{h}_{t}$ in the above formulas (\ref{eq:infer_z})
and (\ref{eq:varphi_z|x}) can be identified with $\boldsymbol{z}_{t}$.

Now, we sum up the discussions thus far as below: 
\begin{itemize}
\item the NNFM (Subsections \ref{subsec:GFM} and \ref{subsec:NNFM})\\
This model consists of the generative rule (\ref{eq:sum_gener_x})
on $\boldsymbol{x}_{t}$, the generative rule (\ref{eq:sum_gener_z})
on $\boldsymbol{z}_{t}$ and the recurrence relation (\ref{eq:sum_reccurence})
for $\boldsymbol{h}_{t}$ in the following forms: 
\begin{gather}
\boldsymbol{x}_{t}|\boldsymbol{z}_{t}\sim\boldsymbol{N}\left(\boldsymbol{\mu}_{x|z,t},\boldsymbol{\sigma}_{x|z,t}^{2}\boldsymbol{I}_{n_{x}}\right),\;{\rm where}\;[\boldsymbol{\mu}_{x|z,t},\boldsymbol{\sigma}_{x|z,t}^{2}]={\rm FNN}_{x|z}\left(\boldsymbol{z}_{t};\Theta_{x|z}\right),\label{eq:sum_gener_x}\\
\boldsymbol{z}_{t}|\boldsymbol{h}_{t-1}\sim\boldsymbol{N}\left(\boldsymbol{\mu}_{z,t},\boldsymbol{\sigma}_{z,t}^{2}\boldsymbol{I}_{n_{z}}\right),\;{\rm where}\;[\boldsymbol{\mu}_{z,t},\boldsymbol{\sigma}_{z,t}^{2}]={\rm FNN}_{z}\left(\boldsymbol{h}_{t-1};\Theta_{z}\right),\label{eq:sum_gener_z}
\end{gather}
and
\begin{gather}
\boldsymbol{h}_{t}={\rm FNN}_{h}\left(\boldsymbol{z}_{t},\boldsymbol{h}_{t-1};\Theta_{h}\right).\label{eq:sum_reccurence}
\end{gather}
In the case of the M-NNFM (a submodel of the NNFM), $\boldsymbol{\mu}_{x|z,t}$
and $\boldsymbol{\sigma}_{x|z,t}$ in (\ref{eq:sum_gener_x}) are
specified by
\begin{gather*}
\boldsymbol{\mu}_{x|z,t}=-{\rm MI\mhyphen FNN}_{x|z,\mu}\left(z_{t,1};\Theta_{x|z}\right),\;\boldsymbol{\sigma}_{x|z,t}^{2}={\rm MI\mhyphen FNN}_{x|z,\sigma}\left(z_{t,1};\Theta_{x|z}\right),
\end{gather*}
or
\begin{gather*}
\boldsymbol{\mu}_{x|z,t}={\rm MI\mhyphen FNN}_{x|z,\mu}\left(z_{t,1};\Theta_{x|z}\right),\;\boldsymbol{\sigma}_{x|z,t}^{2}={\rm MI\mhyphen FNN}_{x|z,\sigma}\left(z_{t,2};\Theta_{x|z}\right),
\end{gather*}
The first and second specifications correspond to the MI-NNFM(1) and
the MI-NNFM(2), respectively.
\item the VINN (Subsection \ref{subsec:VINN})\\
The VINN is designed to approximate the posterior $p(\boldsymbol{z}_{\leq T}|\boldsymbol{x}_{\leq T};\Theta)$
for the NNFM with the conditional normal approximation $q(\boldsymbol{z}_{\leq Teq}|\boldsymbol{x}_{\leq T};\phi)$
as follows:
\begin{gather}
\boldsymbol{z}_{t}|\boldsymbol{x}_{t},\boldsymbol{h}_{t-1}\sim\boldsymbol{N}\left(\boldsymbol{\mu}_{z|x,t},\boldsymbol{\sigma}_{z|x,t}^{2}\boldsymbol{I}_{n_{z}}\right),\;{\rm where}\;[\boldsymbol{\mu}_{z|x,t},\boldsymbol{\sigma}_{z|x,t}^{2}]={\rm FNN}_{z|x}\left(\boldsymbol{x}_{t},\boldsymbol{h}_{t-1};\phi\right),\label{eq:sum_infer_z}
\end{gather}
where $\Theta=\Theta_{x|z}\cup\Theta_{z}\cup\Theta_{h}$. The VINN
can be used for estimations of the parametric factor models in Subsection
\ref{subsec:existing_models}. In this case, $\boldsymbol{h}_{t}$
is replaced by $\boldsymbol{z}_{t}$ in the above formula.
\item the maximizing objective function (Subsection \ref{subsec:VI})\\
Applying the known form for the KL divergence between two normal distributions,
the variational lower bound (\ref{eq:variational_lower_bound}) is
written by
\begin{align*}
\mathcal{L}\left(\Theta,\phi;\boldsymbol{x}_{\leq T}\right) & =\sum_{t=1}^{T}\left(\log p_{q\left(\phi\right)}\left(\left.\boldsymbol{x}_{t}\right|\boldsymbol{z}_{t};\Theta\right)+\mathcal{H}\left(\Theta,\phi\right)\right),
\end{align*}
where
\[
\mathcal{H}\left(\Theta,\phi\right)=\sum_{j=1}^{n_{z}}\frac{1}{2}\left(1-\left(\sigma_{z|x,t}^{\left(j\right)}\middle/\sigma_{z,t}^{\left(j\right)}\right)^{2}-\left(\left(\mu_{z|x,t}^{\left(j\right)}-\mu_{z,t}^{\left(j\right)}\right)\middle/\sigma_{z,t}^{\left(j\right)}\right)^{2}+\log\left(\sigma_{z|x,t}^{\left(j\right)}\middle/\sigma_{z,t}^{\left(j\right)}\right)^{2}\right).
\]
Here, the superscript $j$ indicates the $j$th component of the relevant
quantity, and the subscript $q(\phi)$ clarifies that $\boldsymbol{z}_{t}$
is sampled from the distribution $q$. Note that $(\boldsymbol{\mu}_{z,t},\boldsymbol{\sigma}_{z,t}^{2})$
and $(\boldsymbol{\mu}_{z|x,t},\boldsymbol{\sigma}_{z|x,t}^{2})$
depend on $\Theta$ and $\phi$, respectively. By maximizing $\mathcal{L}(\Theta,\phi;\boldsymbol{x}_{\leq T})$
with respect to $(\Theta,\phi)$, $q(\boldsymbol{z}_{\leq T}|\boldsymbol{x}_{\leq T};\phi)$
and $\Theta$ can be achieved.
\end{itemize}
The FNNs for $\boldsymbol{\sigma}_{t}^{2}$ should have positive activation
functions such as the softplus function in its last layer.

\begin{figure}
\centering{}\includegraphics[scale=0.52]{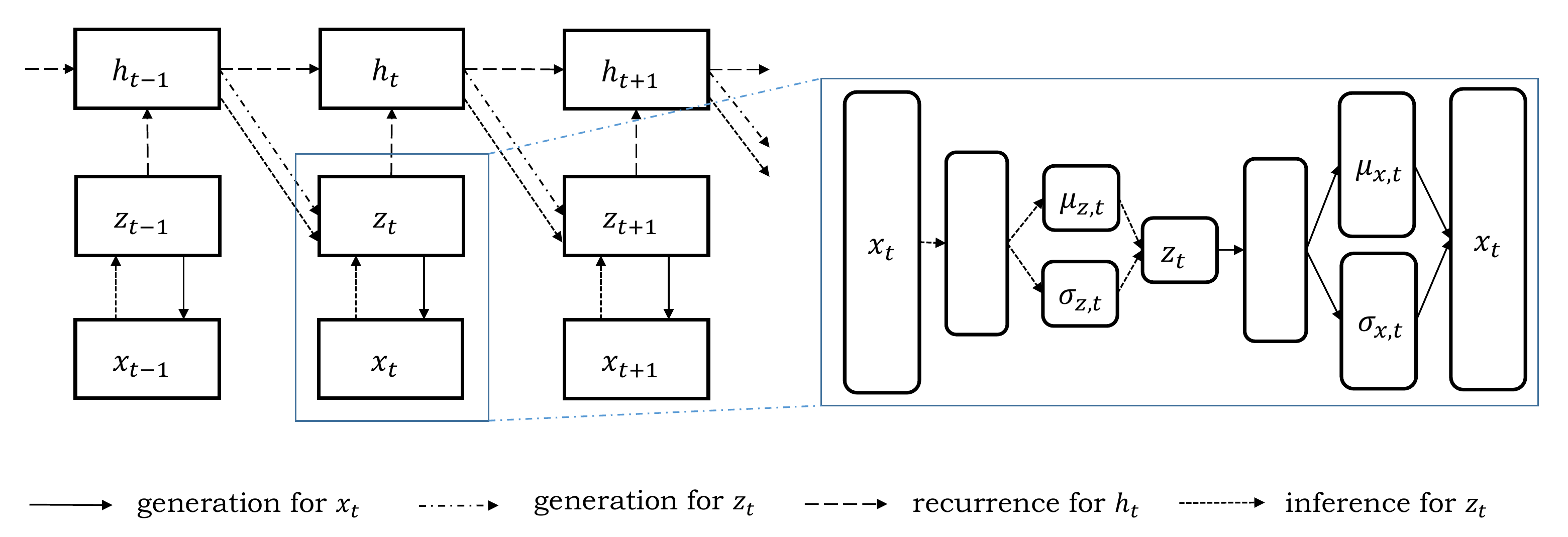}\caption{\label{fig:VRNN}This figure shows the structure of the NNFM and the
VINN. The solid line, the dot-and-dash line, the dotted line and the
dashed line schematize the generative rule (\ref{eq:sum_gener_x})
on $\boldsymbol{x}_{t}$, the generative rule (\ref{eq:sum_gener_z})
on $\boldsymbol{z}_{t}$, the recurrence relation (\ref{eq:sum_reccurence})
for $\boldsymbol{h}_{t}$ and the variational approximation (\ref{eq:sum_infer_z})
on $\boldsymbol{z}_{t}$, respectively. The area in blue lines is
described in more detail on its right side, which represents the interrelation
between $\boldsymbol{x}_{t}$ and $\boldsymbol{z}_{t}$ at a fixed
time.}
\end{figure}

Figure \ref{fig:VRNN} shows the structure of the NNFM and the VINN
in which the directed arrows schematize the generative rule (\ref{eq:sum_gener_x})
on $\boldsymbol{x}_{t}$, the generative rule (\ref{eq:sum_gener_z})
on $\boldsymbol{z}_{t}$, the recurrence relation (\ref{eq:sum_reccurence})
for $\boldsymbol{h}_{t}$ and the variational approximation (\ref{eq:sum_infer_z})
on $\boldsymbol{z}_{t}$. Particularly, observe the part of the diagram
in blue lines, which is described in more detail on its right side.
The enlarged figure represents the interrelation between $\boldsymbol{x}_{t}$
and $\boldsymbol{z}_{t}$ at a fixed time. When judging only by this
part, it is not different from the variational autoencoder. Autoencoders,
including the variational autoencoder, aim to learn a representation
for a set of data, typically for the purpose of dimensionality reduction.
To this end, an autoencoder compresses a set of data into a short
code, and then uncompresses it so that the output of the network closely
matches the original data. So, the deepest layer for $\boldsymbol{z}_{t}$
should be narrow in order to make the layer a bottleneck, as one can
check through the figure. By doing so, we can leave essential features
of the given data to the layer. In our work, we insert daily asset
returns of many companies into the bottleneck network and seek suitable
factors for systematic risk. Similar ideas are often shown in different
fields (cf. \citet{mikolov2013efficient,heaton2017deep}), but the
papers, if any, which associate the idea with measuring systematic
risk, to the best of author's knowledge, have yet to appear.

Stochastic neural networks can not be easily trained by a naive back-propagation
for deterministic networks such as the FNN and the RNN. The key concept
to resolve the problem is a so-called reparametrization trick devised
in \citet{kingma2013auto}, which has already been reflected in the
design process of the NNFM.

\section{\label{sec:empirical_analysis}Empirical Analysis}

In this section, we compare the NNFM with the parametric factor models
in Subsection \ref{subsec:existing_models} in terms of performance
for measuring systematic risk. We denote the models by MODEL\_NAME$(n_{z})$,
and $n_{z}$, the number of systematic factors, is set no more than
2 so as to induce the intuitive relation between $\boldsymbol{z}_{t}$
and the risk. The following 12 models are on the table: 8 parametric
factor models (APT(1), L-SVFM(1), SR-SVFM(1), APT(2), L-SVFM(2), SR-SVFM(2),
APT-L(2), APT-SR(2)) and 4 network-based models (NNFM(1), NNFM(2),
M-NNFM(1), M-NNFM(2)).

As mentioned, we wish to find out systematic factors through daily
returns on many assets. To do this, we collect daily returns of 81$(=n_{x})$
companies among the components of S\&P 100, which consists of 5,240-day
data from 16 May 1997 to 13 March 2018. We then number the dates in
the period from $t=1$ to $t=T$. On the basis of the dividing date
$T_{d}$ (10 August 2011), which is chosen among the European debt
crisis, the data $\boldsymbol{x}_{\leq T}$ is divided into the training
data $\boldsymbol{x}_{\leq T}^{\left(1\right)}(:=\{\boldsymbol{x}_{1},\cdots,\boldsymbol{x}_{T_{d-1}}\})$
with the first 3,582-day data (68.4\%) and the test data $\boldsymbol{x}_{\leq T}^{\left(2\right)}(:=\{\boldsymbol{x}_{T_{d}},\cdots,\boldsymbol{x}_{T}\})$
with the last 1,658-day data (31.6\%). The 19 components are counted
out because their stocks were listed too recently to provide enough
data. See the table in \ref{sec: S=000026P list} for the company
list in which their symbols and sectors are summarized. 

\begin{figure}
\centering{}\includegraphics[scale=0.7]{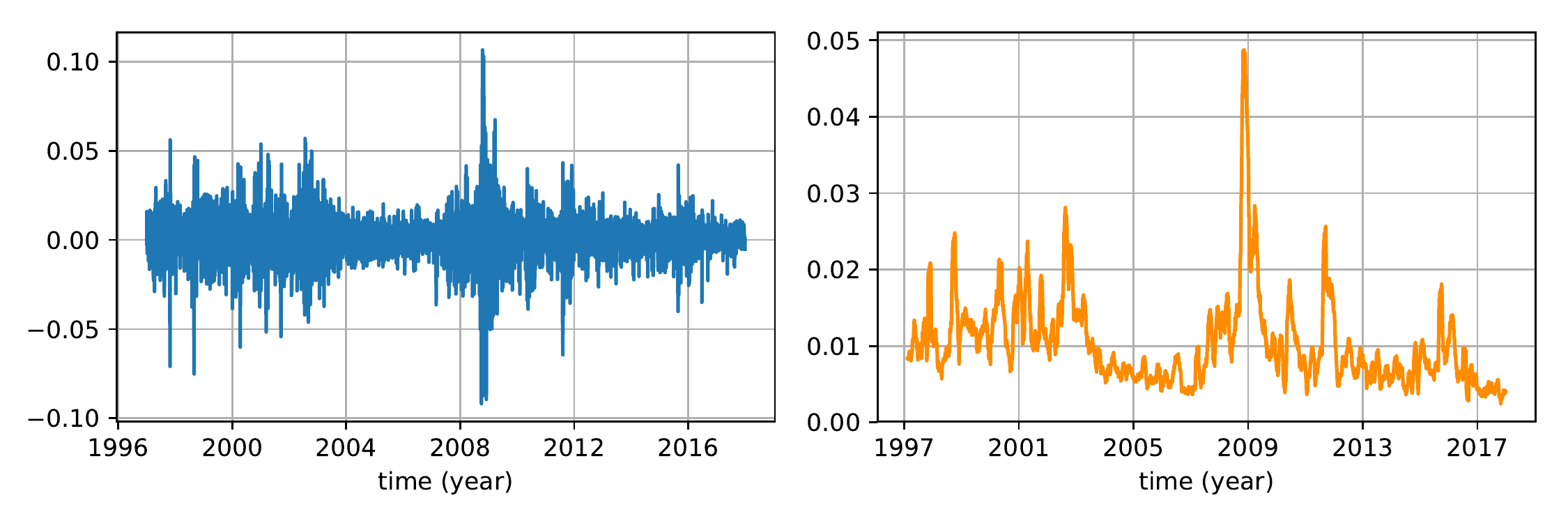}\caption{\label{fig:OEX} This figure displays the daily returns of the S\&P
100 (left) and its 30-day historical volatilities (right) during the
data period.}
\end{figure}

\begin{table}
\begin{centering}
\subfloat[\label{tab:oex_statistics}statistics for daily returns of S\&P 100]{\centering{}%
\begin{tabular}{ccccccccc}
\toprule 
Min & Q1 & Q2 & Q3 & Max & M1 & M2 & M3 & M4\tabularnewline
\midrule 
-9.19e-2 & -5.11e-3 & 5.50e-4 & 5.91e-03 & 1.07e-1 & 2.25e-4 & 1.21e-2 & -1.88e-1 & 7.48e+0\tabularnewline
\bottomrule
\end{tabular} }
\par\end{centering}
\centering{}\subfloat[\label{tab:componets_statistics}statistics for daily returns of S\&P
100 components]{\centering{}%
\begin{tabular}{cccccccccc}
\toprule 
 & Min & Q1 & Q2 & Q3 & Max & M1 & M2 & M3 & M4\tabularnewline
\midrule
Min & -4.07e-1 & -1.25e-1 & -8.34e-2 & -6.46e-2 & -3.84e-2 & -1.04e-1 & 6.34e-2 & -2.27e+0 & 6.29e+0\tabularnewline
Q1 & -6.07e-3 & -4.35e-3 & -3.77e-3 & -3.28e-3 & -2.45e-3 & -3.90e-3 & 8.33e-4 & -4.96e-1 & -1.67e-1\tabularnewline
Q2 & -3.35e-5 & 7.68e-5 & 1.44e-4 & 2.10e-4 & 3.89e-4 & 1.44e-4 & 1.05e-4 & 2.61e-1 & -5.42e-1\tabularnewline
Q3 & 2.79e-3 & 3.71e-3 & 4.14e-3 & 4.71e-3 & 6.83e-3 & 4.30e-3 & 8.86e-4 & 6.93e-1 & 3.34e-1\tabularnewline
Max & 3.76e-2 & 5.94e-2 & 7.20e-2 & 9.51e-2 & 2.72e-1 & 8.25e-2 & 3.93e-2 & 2.37e+0 & 7.12e+0\tabularnewline
 &  &  &  &  &  &  &  &  & \tabularnewline
M1 & -1.79e-4 & 1.32e-4 & 1.75e-4 & 2.26e-4 & 5.55e-4 & 1.85e-4 & 1.02e-4 & 6.36e-1 & 4.00e+0\tabularnewline
M2 & 5.27e-3 & 7.36e-3 & 8.69e-3 & 1.03e-2 & 1.65e-2 & 9.12e-3 & 2.41e-3 & 9.56e-1 & 5.44e-1\tabularnewline
M3 & -4.93e+0 & -3.54e-1 & -1.14e-1 & 6.54e-2 & 2.62e+0 & -3.33e-1 & 9.66e-1 &  -1.97e+0 & 7.56e+0\tabularnewline
M4 & 3.78e+0 & 6.89e+0 & 8.94e+0 & 1.67e+1 & 1.51e+2 & 1.78e+1 & 2.42e+1 & 3.55e+0 & 1.39e+1\tabularnewline
\bottomrule
\end{tabular} }\caption{\label{tab:statistics}These tables list the statistics for daily
returns of S\&P 100 (upper) and its components (lower). The lower
table shows the statistics obtained by processing the statistics for
the components again. For example, the skewness of the worst returns
for the components is -2.27. The abbreviation M$n$ stands for the
$n$-th moment; expectation (M1), deviation (M2), skewness (M3) and
kurtosis (M4). The kurtosis here is based upon Fisher’s definition,
that is to say, it gives zero for a normal distribution.}
\end{table}

Figure \ref{fig:OEX} displays the daily returns of the S\&P 100 (left)
and its 30-day historical volatilities (right) during the data period.
As one can see, several crises influenced the US market; the Asian
financial crisis (1997-1998), the dot-com bubble (2000-2002), the
mortgage crisis (2008-2009), the European debt crisis (2010-2011)
and the Chinese stock market turbulence (2015-2016). Based on the
30-day volatilities, the S\&P 100 was affected the most by the mortgage
crisis. Table \ref{tab:statistics} lists the statistics for daily
returns of S\&P 100 (upper) and its components (lower). The lower
table displays the statistics obtained by processing the statistics
for the components again. For example, the value -2.27 in the table
indicates the skewness of the worst returns for the components. The
upper table shows typical values for an unconditional distribution
of index returns; slight negative skewness and strong kurtosis. On
the contrary, the statistical properties of the components' returns
can not be encapsulated into a few words. For instance, the returns
for blue-chip companies are positively skewed, while less successful
companies have negatively skewed returns. Because the distributions
are distinct from one another, one kind of model may be difficult
to cover the dynamics of all the components (for example, a belief
that all assets follow the GARCH(1,1) seems naive for us). This implies
the necessity of nonparametric models such as the NNFM.

We infer the approximate posteriors $q(\boldsymbol{z}_{\leq T}|\boldsymbol{x}_{\leq T};\phi)$
and the parameters $\Theta$ for the models with the VI. To be concrete,
we find $\Theta$ and $\phi$ by maximizing the variational lower
bound $\mathcal{L}(\Theta,\phi;\boldsymbol{x}_{\leq T}^{\left(1\right)})$
for the training data $\boldsymbol{x}_{\leq T}^{\left(1\right)}$
and subsequently compute $\mathcal{L}(\Theta,\phi;\boldsymbol{x}_{\leq T}^{\left(2\right)})$
for the test data $\boldsymbol{x}_{\leq T}^{\left(2\right)}$. This
is implemented through Tensorflow\footnote{https://www.tensorflow.org},
the deep-learning framework developed by Google. In the learning
process, we put $n_{h}=3n_{z}$ and also let each FNN for the NNFM
have one hidden layer. In addition, the ADAM optimizer (\citet{kingma2014adam})
is used to find a good local extremum of $\mathcal{L}(\Theta,\phi;\boldsymbol{x}_{\leq T}^{\left(1\right)})$.
The learning iterations are repeated during 2,000 epochs with mini-batches
of size 50 (refer to \citet{goodfellow2016deep} for terminologies
such as the mini-batch and the epoch).

\begin{figure}
\centering{}\includegraphics[scale=0.7]{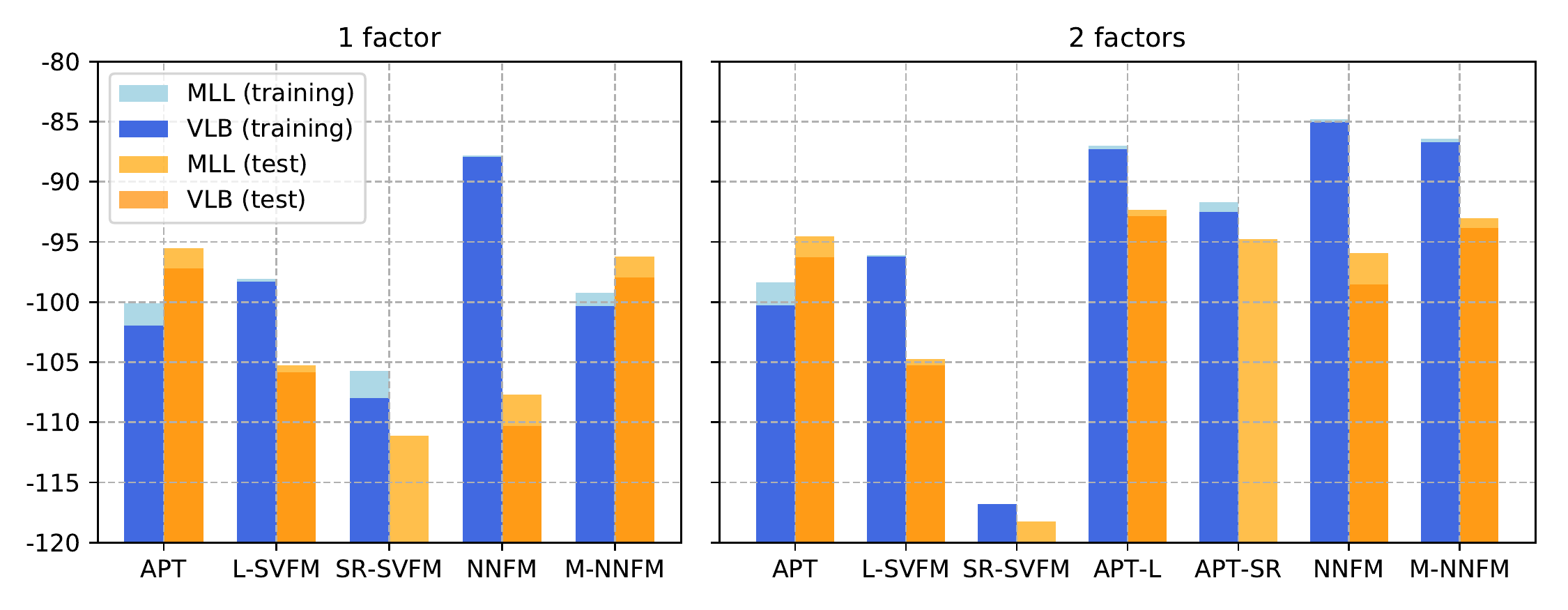}\caption{\label{fig:MLL_VLB}This figure shows the approximate marginal log-likelihood
(MLL) and the variational lower bound (VLB) for the factor models
for the training data $\boldsymbol{x}_{\protect\leq T}^{\left(1\right)}$
and the test data $\boldsymbol{x}_{\protect\leq T}^{\left(2\right)}$.
The quantities graphed in the figure are the values of the two criteria
divided by the length of each data. Here, the scores of the SR-SVFM(2)
are indicated just for reference. They are not meaningful because
the model is underfit. }
\end{figure}

In what follows, we evaluate performance of the models for measuring
systematic risk with the two criteria for the VI: the marginal log-likelihood
(MLL) $\log p_{q\left(\phi\right)}(\boldsymbol{x}_{\leq T}^{\left(k\right)};\Theta)$
and the variational lower bound (VLB) $\mathcal{L}(\Theta,\phi;\boldsymbol{x}_{\leq T}^{\left(k\right)})$
for $k=1,2$. Because $\log p_{q\left(\phi\right)}(\boldsymbol{x}_{\leq T}^{\left(k\right)};\Theta)$
is intractable for most specifications of the GFM, the following simulation-based
approximation is applied: 
\begin{align*}
\log p_{q\left(\phi\right)}\left(\boldsymbol{x}_{\le T}^{\left(k\right)};\Theta\right) & \approx\frac{1}{L}\sum_{l=1}^{L}\left(\log p_{q\left(\phi\right)}\left(\boldsymbol{x}_{\le T}^{\left(k\right)}|\boldsymbol{z}_{\le T}^{\left(l\right)};\Theta\right)+\log p\left(\boldsymbol{z}_{\le T}^{\left(l\right)};\Theta\right)-\log q\left(\boldsymbol{z}_{\le T}^{\left(l\right)}|\boldsymbol{x}_{\le T}^{\left(k\right)};\phi\right)\right),
\end{align*}
where the number of simulations, denoted by $L$, should be large
enough. This method is effective when $n_{z}<5$ (\citet{kingma2013auto}).
Figure \ref{fig:MLL_VLB} shows $\log p_{q\left(\phi\right)}(\boldsymbol{x}_{\leq T}^{\left(k\right)};\Theta)$
and $\mathcal{L}(\Theta,\phi;\boldsymbol{x}_{\leq T}^{\left(k\right)})$
for the models. The quantities graphed in the figure are the values
of the two criteria divided by the length of each data. Note that
the VLB bar is not lower than the MLL bar for every model, because
$\log p_{q\left(\phi\right)}(\boldsymbol{x}_{\leq T}^{\left(k\right)};\Theta)\geq\mathcal{L}(\Theta,\phi;\boldsymbol{x}_{\leq T}^{\left(k\right)})$
always holds. Meanwhile, we list the point estimates for the parametric
factor models in AppendixB. Regarding the parameters associated with
all companies such as $\alpha_{0,i}$, their averages and deviations
are provided instead.

For the one-factor models, the APT(1) and the M-NNFM(1) outperform
the other one-factor models. Although the NNFM(1) is outstanding for
$\boldsymbol{x}_{\leq T}^{\left(1\right)}$, its performance is somewhat
disappointing for $\boldsymbol{x}_{\leq T}^{\left(2\right)}$. We
guess that the NNFM(1) may be overfit because any financial knowledge
is not accommodated into the model. Under the NNFM(1), the components
of $\boldsymbol{\mu}_{x|z,t}$ and $\boldsymbol{\sigma}_{x|z,t}$
in the generative rule (\ref{eq:generative_x}) for $\boldsymbol{x_{t}}$
can move in the direction irrelevant to that of the systematic factor
$\boldsymbol{z}_{t}$. But the events seem to occur rarely in practice.
Based on the test result, the M-NNFM(1) is a better choice to prevent
the overfitting. The L-SVFM(1) and the SR-SVFM(1) also produce inferior
outcomes for $\boldsymbol{x}_{\leq T}^{\left(2\right)}$ to those
for $\boldsymbol{x}_{\leq T}^{\left(1\right)}$, though to a lesser
extent than the NNFM(1). It may make the difference that asset prices
for the test period are much less volatile than those for the training
period. In addition, note that the SR-SVFM(1) is insufficient for
providing satisfactory explanations compared with the L-SVFM(1). This
agrees with several comparative researches that logarithmic form of
volatility is more suitable to market data than square-root form.

With regard to the two-factor models, the APT-L(2) and the M-NNFM(2)
show great fitting capacities for both the $\boldsymbol{x}_{\leq T}^{\left(1\right)}$
and $\boldsymbol{x}_{\leq T}^{\left(2\right)}$. The NNFM(2) still
seems to suffer a little from the overfitting problem. The results
are relatively unsatisfactory for the models APT(2), L-SVFM(2), SR-SVFM(2),
in which the two factors are all associated with exactly one of $\boldsymbol{\mu}_{x|z,t}$
and $\boldsymbol{\sigma}_{x|z,t}$.  The most effective models, the
APT-L(2) and the M-NNFM(2), have respective factors for $\boldsymbol{\mu}_{x|z,t}$
and $\boldsymbol{\sigma}_{x|z,t}$. On the other hand, one can realize
that the fitting scores for the SR-SVFM(2) are fairly low. In fact,
we can not detect a good local maximum for the model. Namely, we speculate
that the SR-SVFM(2) is underfit and its results are rather not meaningful.
The bar graphs for the model are drawn just for reference. It tells
us that the affine models with several factors for $\boldsymbol{\sigma}_{x|z,t}$
may be difficult to deal with. Furthermore, the affine models, the
SR-SVFM(2) and the APT-SR(2), give the worse results than the logarithmic
models, the L-SVFM(2) and the APT-L(2), as in the case of the one-factor
models. 
\begin{figure}
\centering{}\includegraphics[scale=0.7]{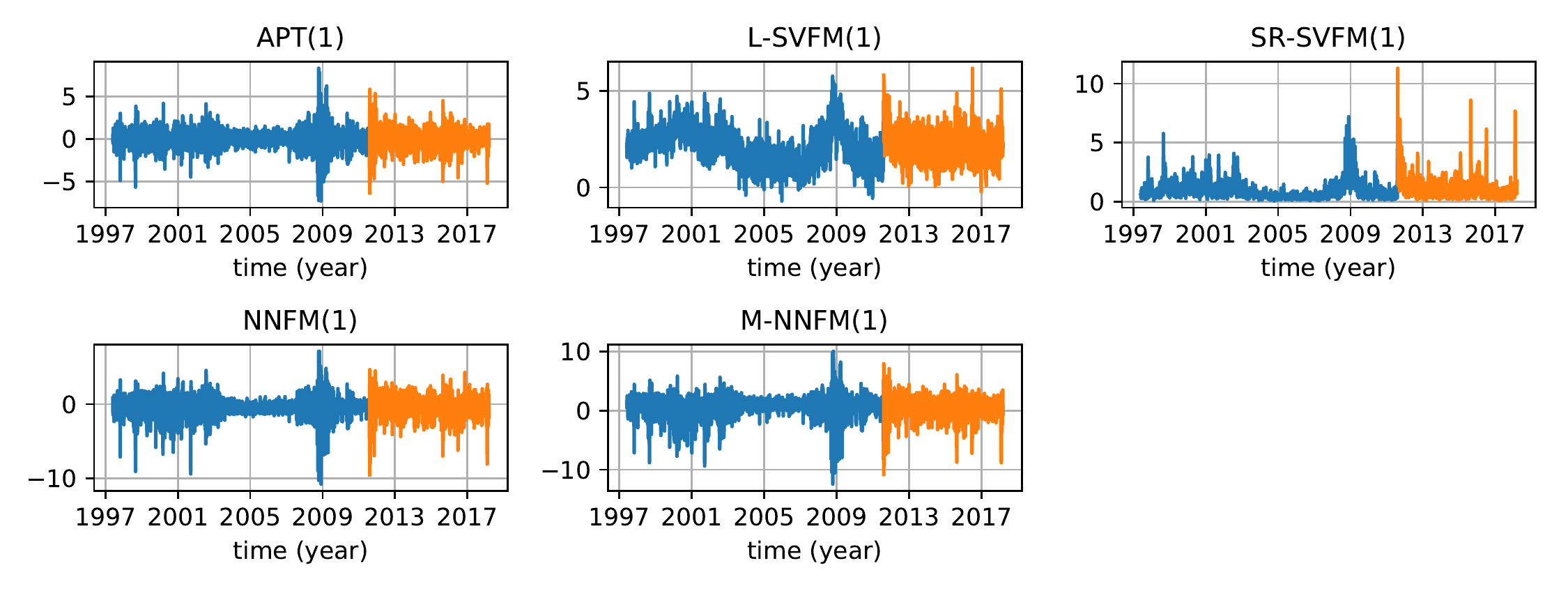}\caption{\label{fig:1_factor_z}This depicts the factor process $\boldsymbol{z}_{t}=(z_{t,1})$
for the one-factor models. The blue and orange colors are used to
distinguish between the training period and the test period.}
\end{figure}
\begin{figure}
\begin{centering}
\subfloat[the first factor $z_{t,1}$]{\centering{}\includegraphics[scale=0.7]{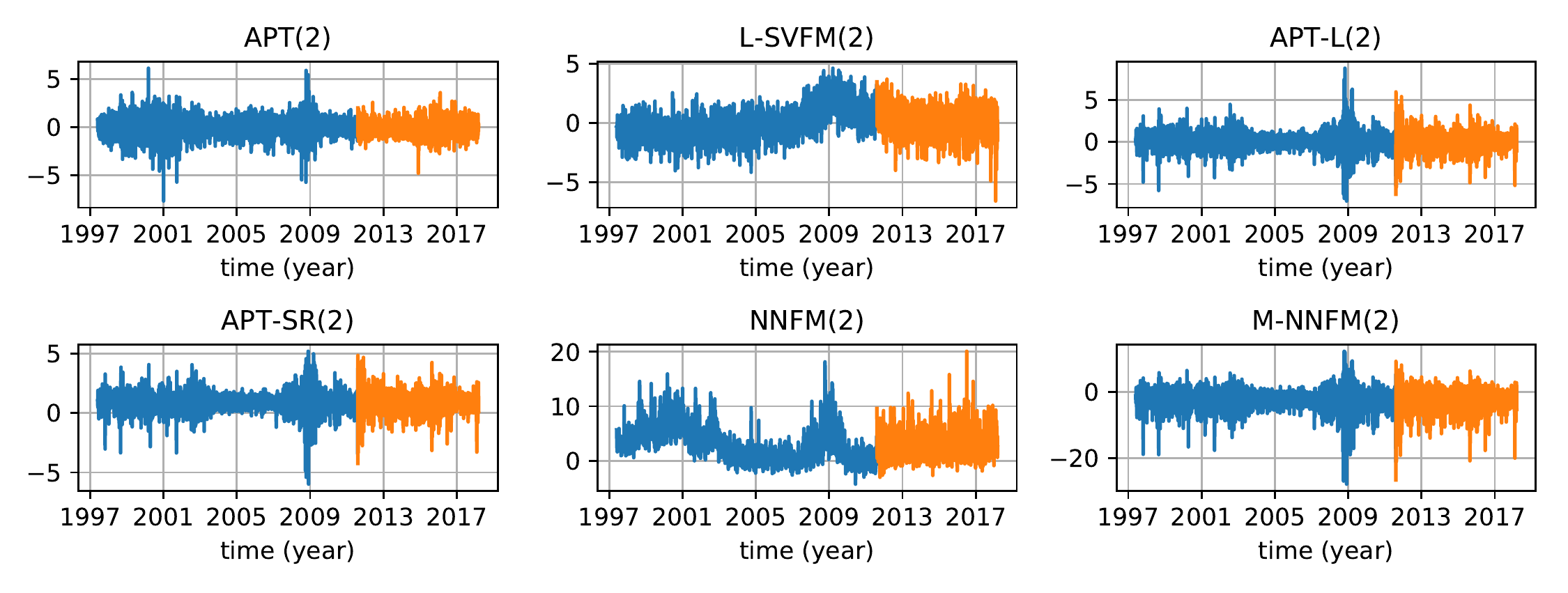}}
\par\end{centering}
\centering{}\subfloat[the second factor $z_{t,2}$]{\centering{}\includegraphics[scale=0.7]{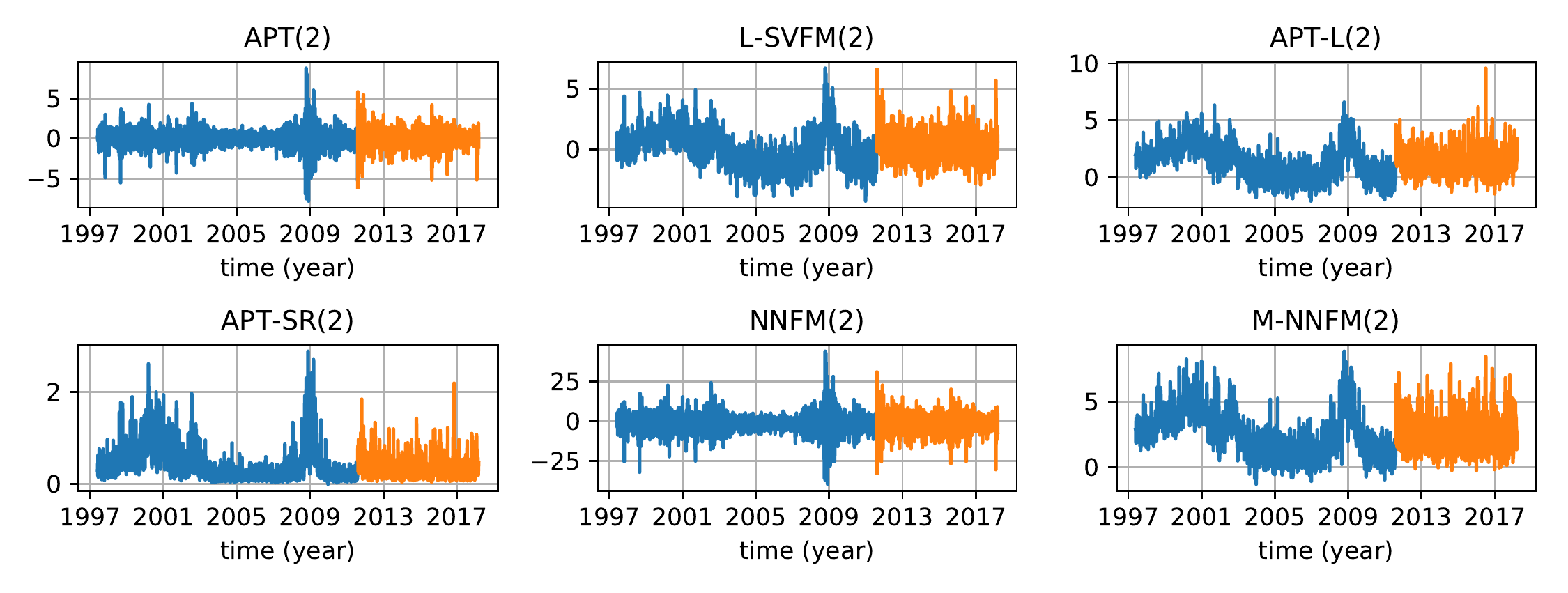}}\caption{\label{fig:2_factor_z}This draws the factor process $\boldsymbol{z}_{t}=(z_{t,1},z_{t,2})$
for the two-factor models. The graphs for the SR-SVFM(2) are left
out of consideration because the model is underfit. The blue and orange
lines mean the training period and the test period, respectively.}
\end{figure}

Figures \ref{fig:1_factor_z} and \ref{fig:2_factor_z} depict $\boldsymbol{z}_{t}=(z_{t,1})$
for the one-factor models and $\boldsymbol{z}_{t}=(z_{t,1},z_{t,2})$
for the two-factor models, respectively. We prescind from the graphs
for the SR-SVFM(2) because the model is underfit. The blue and orange
colors are used to distinguish between the training period and the
test period. Interestingly, $\boldsymbol{z}_{t}$ for the one-factor
network-based models, the NNFM(1) and the M-NNFM(1), behave like $\boldsymbol{z}_{t}$
for the APT(1). Moreover, the trajectories of $\boldsymbol{z}_{t}$
for the two-factor network-based models, the NNFM(2) and the M-NNFM(2),
also look like those of $\boldsymbol{z}_{t}$ for the APT-L(2). Perhaps,
this is because the APT(1) and the APT-L(2) are highly suitable to
the data in the respective classes of the one-factor models and the
two-factor models.

The discussions so far deduce several consequences. Above all things,
the GFM should have at least one factor for $\boldsymbol{\mu}_{x|z,t}$.
Otherwise, $q(\boldsymbol{z}_{\leq T}|\boldsymbol{x}_{\leq T};\phi)$
and $\Theta$ may be biased, so it can impair fitting performance
for the test data as in the cases of the SVFMs. Provided that the
condition is satisfied, it is desirable to take into consideration
one additional factor for $\boldsymbol{\sigma}_{x|z,t}$. Then, the
fitting score for $\boldsymbol{x}_{\leq T}^{\left(1\right)}$ is improved
greatly, and the score for $\boldsymbol{x}_{\leq T}^{\left(2\right)}$
is increased modestly. But one should not jump to a conclusion that
the additional factor generally yields a bigger improvement on training
data than on test data. Note that financial time series is usually
inhomogeneous. In addition, the M-NNFM is more appropriate than the
NNFM in terms of stability. The NNFM produces good results for $\boldsymbol{x}_{\leq T}^{\left(1\right)}$
but does not for $\boldsymbol{x}_{\leq T}^{\left(2\right)}.$ Lastly,
speaking only with our test, the network-based models can not deliver
better performance than the best ones among the existing parametric
models. In our view, it arises from the limitation of the VI. As said
before, the VI is a method to approximate the posterior $p(\boldsymbol{z}_{\leq T}|\boldsymbol{x}_{\leq T};\Theta)$
as a normal distribution $q(\boldsymbol{z}_{\leq T}|\boldsymbol{x}_{\leq T};\phi)$.
We might be unable to draw the maximum capacities of the models because
the VI gives the approximate objective function $\mathcal{L}(\Theta,\phi;\boldsymbol{x}_{\leq T})$,
which is flatter and shallower than the true objective function. 

Although we can not derive overwhelming superiority of the network-based
models to the parametric factor models, the new models are still noteworthy
in that they do not require any prior knowledge. A researcher using
the models (not NNFM but M-NNFM, to be precise) need not worry about
which one is a proper choice among the parametric models. Whoever
uses the model can achieve the performance as best the parametric
models could. Furthermore, in order to find estimates more accurately,
alternative methods which do not rely on approximate techniques can
be employed such as generative stochastic networks (\citet{bengio2014deep})
and generative adversarial networks (\citet{goodfellow2014generative}).
These approaches can clarify the differences between the models showing
similar performances in this paper. We leave the topic as further
studies.

\section{Conclusion}

Measuring systematic risk is one of essential works to both financial
supervisory authority and individual companies. In literature, this
work is generally done by one or more common factors. We measure systematic
risk with a new nonparametric factor model, the neural network factor
model. Artificial neural networks are more suitable to modeling the
risk than the parametric factor models holding fixed market views
because they can adapt to various markets flexibly. The proper systematic
factors can be naturally induced by putting long-term returns on a
great many assets into the new network. This is possible because its
deepest layer is designed as a bottleneck in data transmission to
get essential features of the whole stock market.

The neural network factor model is compared with the parametric factor
models using 20 years data of S\&P 100 components. For model comparisons,
two criteria are adopted: marginal log-likelihood and variational
lower bound. Because the first proposed model ends up facing overfitting
issues, we additionally devise the monotone neural network factor
model by restricting the freedom degree of the first model. However,
we can not derive the conclusion that the networks outperform the
best ones among the parametric factor models. This is because we can
not reach the maximum capacities of the models since the variational
inference, the estimation method used for this study, has the limitation
as an approximate approach. Nevertheless, the new models are still
noteworthy in that it achieves the performance as best the comparing
models could without any prior knowledge.

We lastly suggest some subjects for further studies. Because our method
is not fully Bayesian (that is to say, the parameters of the models
are not random), one can repeat the test of this paper using the full
variational inference (cf. \citet{kingma2015variational,gal2016theoretically}).
Non-approximate methods can also be introduced for the test such as
generative stochastic networks (\citet{bengio2014deep}) and generative
adversarial networks (\citet{goodfellow2014generative}). 

\section*{Funding}

This research did not receive any specific grant from funding agencies
in the public, commercial, or not-for-profit sectors.

\section*{Bibliography}

\bibliographystyle{elsarticle-harv}
\bibliography{ref}
\pagebreak{}

\appendix

\section{The S\&P 100 component list\label{sec: S=000026P list}\medskip{}
}
\begin{center}
\begin{tabular}{cccccccccccc}
\toprule 
Symbol & Sector &  & Symbol & Sector &  & Symbol & Sector &  & Symbol & Sector & \tabularnewline
\midrule
\midrule 
AAPL & T. &  & COP & E. &  & INTC & T. &  & PFE & H. & \tabularnewline
\midrule 
ABBV & H. & x & COST & C.D. &  & JNJ & H. &  & PG & C.D. & \tabularnewline
\midrule 
ABT & H. &  & CSCO & T. &  & JPM & F.S. &  & PM & C.D. & x\tabularnewline
\midrule 
ACN & T. & x & CVS & H. &  & KHC & C.D. & x & PYPL & F.S. & x\tabularnewline
\midrule 
AGN & H. &  & CVX & E. &  & KMI & E. & x & QCOM & T. & \tabularnewline
\midrule 
AIG & F.S. &  & DHR & H. &  & KO & C.D. &  & RTN & I. & \tabularnewline
\midrule 
ALL & F.S. &  & DIS & C.C. &  & LLY & H. &  & SBUX  & C.C. & \tabularnewline
\midrule 
AMGN & H. &  & DUK & U. &  & LMT & I. &  & SLB & E. & \tabularnewline
\midrule 
AMZN & C.C. &  & DWDP & B.M. &  & LOW & C.C. &  & SO & U. & \tabularnewline
\midrule 
AXP & F.S. &  & EMR & I. &  & MA & F.S. & x & SPG & R.E. & \tabularnewline
\midrule 
BA  & I. &  & EXC & U. &  & MCD & C.C. &  & T & C.S. & \tabularnewline
\midrule 
BAC & F.S. &  & F & C.C. &  & MDLZ & C.D. & x & TGT & C.D. & \tabularnewline
\midrule 
BIIB & H. &  & FB & T. & x & MDT & H. &  & TWX & C.C. & \tabularnewline
\midrule 
BK & F.S. &  & FDX & I. &  & MET & F.S. & x & TXN & T. & \tabularnewline
\midrule 
BKNG & C.C. & x & FOX & C.C. &  & MMM & I. &  & UNH & H. & \tabularnewline
\midrule 
BLK & F.S. & x & GD & I. &  & MO & C.D. &  & UNP & I. & \tabularnewline
\midrule 
BMY & H. &  & GE & I. &  & MON & B.M. & x & UPS & I. & x\tabularnewline
\midrule 
BRKB & F.S. &  & GILD & H. &  & MRK & H. &  & USB & F.S. & \tabularnewline
\midrule 
C & F.S. &  & GM & C.C. & x & MS & F.S. &  & UTX & I. & \tabularnewline
\midrule 
CAT & I. &  & GOOG & T. & x & MSFT & T. &  & V & F.S. & x\tabularnewline
\midrule 
CELG & H. &  & GS & F.S. & x & NEE & U. &  & VZ & C.S. & \tabularnewline
\midrule 
CHTR & C.S. & x & HAL & E. &  & NKE & C.G. &  & WBA & C.D. & \tabularnewline
\midrule 
CL & C.D. &  & HD & C.C. &  & ORCL & T. &  & WFC & F.S. & \tabularnewline
\midrule 
CMCSA & C.S. &  & HON & I. &  & OXY & E. &  & WMT & C.D. & \tabularnewline
\midrule 
COF & F.S. &  & IBM & T. &  & PEP & C.D. &  & XOM & E. & \tabularnewline
\bottomrule
\end{tabular}
\par\end{center}

\medskip{}

\begin{center}
\begin{tabular}{ccccccccc}
\toprule 
F.S. & Financial Services & 18 & H. & Healthcare & 16 & I. & Industrials & 13\tabularnewline
C.D. & Consumer Defensive & 12 & C.C. & Consumer Cyclical  & 11 & T. & Technology & 11\tabularnewline
E. & Energy & 7 & C.S. & Communication Services & 4 & U. & Utilities & 4\tabularnewline
B.M. & Basic Materials & 2 & R.E. & Real Estate & 1 & C.G. & Consumer Goods & 1\tabularnewline
\bottomrule
\end{tabular}
\par\end{center}

This table shows the company list for the test in which their symbols
and sectors are summarized. The full names of the companies can be
quickly restored by a search engine. The 19 ones among the companies
are excluded because their stocks were listed too recently to provide
enough data (checked by `x').

\pagebreak{}

\begin{sidewaystable}

\section{Point estimates for the parametric factor models}
\begin{centering}
\begin{tabular}{ccccccccccc}
\toprule 
 & $\alpha_{0,i}$ & $\alpha_{1,i}$ & $\alpha_{2,i}$ & $\beta_{0,i}$ & $\beta_{1,i}$ & $\beta_{2,i}$ & $c_{0}$ & $c_{1}$ & $a_{0}$ & $a_{1}$\tabularnewline
\midrule
APT(1) & 4.414e-4 & 5.509e-1 &  & 8.237e-1 &  &  &  &  &  & \tabularnewline
 & (1.689e-3) & (1.050e-1) &  & (7.347e-2) &  &  &  &  &  & \tabularnewline
 &  &  &  &  &  &  &  &  &  & \tabularnewline
L-SVFM(1) & 1.250e-2 &  &  & -1.301e+0 & 5.270e-1 &  &  &  & 9.556e-1 & \tabularnewline
 & (1.146e-2) &  &  & (2.679e-1) & (9.002e-2) &  &  &  &  & \tabularnewline
 &  &  &  &  &  &  &  &  &  & \tabularnewline
SR-SVFM(1) & 1.664e-2 &  &  & 5.593e-2 & 3.630e-1 &  & 3.204e0 &  & 4.984e-1 & \tabularnewline
 & (1.016e-2) &  &  & (6.604e-2) & (3.492e-2) &  &  &  &  & \tabularnewline
 &  &  &  &  &  &  &  &  &  & \tabularnewline
APT(2) & 9.190e-4 & 5.495e-1 & 1.377e-1 & 7.994e-1 &  &  &  &  &  & \tabularnewline
 & (1.503e-3) & (1.048e-1) & (1.280e-1) & (9.880e-2) &  &  &  &  &  & \tabularnewline
 &  &  &  &  &  &  &  &  &  & \tabularnewline
APT-L(2) & -2.125e-3 & 5.4521e-1 &  & -7.782e-1 & 3.080e-1 &  &  &  & 9.100e-1 & \tabularnewline
 & (1.044e-2) & (8.888e-2) &  & (1.811e-1) & (5.668e-2) &  &  &  &  & \tabularnewline
 &  &  &  &  &  &  &  &  &  & \tabularnewline
APT-SR(2) & 7.987e-1 & 1.009e0 &  & 3.533e-2 & 9.585e-1 &  & 9.424e-1 &  & 7.012e-1 & \tabularnewline
 & (1.320e-1) & (1.687e-1) &  & (5.808e-2) & (1.910e-1) &  &  &  &  & \tabularnewline
\bottomrule
\end{tabular}
\par\end{centering}
\bigskip{}

\centering{}
\begin{gather*}
x_{t,i}=\alpha_{0,i}+\alpha_{1,i}z_{t,1}^{\left(1\right)}+\alpha_{2,i}z_{t,2}^{\left(1\right)}+f\left(\beta_{0,i}+\beta_{1,i}z_{t,1}^{\left(2\right)}+\beta_{2,i}z_{t,2}^{\left(2\right)}\right)\epsilon_{t,i},\quad\epsilon_{t,i}\sim N\left(0,1\right)\\
z_{t,j}^{\left(1\right)}=e_{t,j},\quad z_{t,j}^{\left(2\right)}=c_{j}+a_{j}z_{t-1,j}^{\left(2\right)}+e_{t,j},\quad e_{t,j}\sim N\left(0,1\right)
\end{gather*}
\\
APT: $\beta_{1,i}=\beta_{2,i}=0$, $f\left(x\right)=x$; \\
L-SVFM: $\alpha_{1,i}=\alpha_{2,i}=c_{1}=c_{2}=0$, $f\left(x\right)=e^{x}$;
SR-SVFM: $\alpha_{1,i}=\alpha_{2,i}=0$, $f\left(x\right)=\sqrt{x}$;
\\
APT-L: $\alpha_{2,i}=\beta_{2,i}=c_{1}=0$, $f\left(x\right)=e^{x}$;
APT-SR: $\alpha_{2,i}=\beta_{2,i}=0$, $f\left(x\right)=\sqrt{x}$
\caption{We list the point estimates for the parametric factor models in Subsection
\ref{subsec:existing_models}. Regarding the parameters associated
with all companies such as $\alpha_{0,i}$, the averages and the deviations
(in parenthesis) of them are provided instead. The values for the
SR-SVFM(2) are left out of consideration because the model is underfit.}
\end{sidewaystable}

\end{document}